# Exact solutions of Dirac and Schrödinger equations for a large class of power-law potentials at zero energy

**A. D. Alhaidari**

Physics Department, King Fahd University of Petroleum & Minerals, Box 5047, Dhahran 31261, Saudi Arabia

E-mail: haidari@mailaps.org

We obtain exact solutions of Dirac equation at zero kinetic energy for radial power-law relativistic potentials. It turns out that these are the relativistic extension of a subclass of exact solutions of Schrödinger equation with two-term power-law potentials at zero energy. The latter is solved by point canonical transformation of the exactly solvable problem of the three dimensional oscillator. The wavefunction solutions are written in terms of the confluent hypergeometric functions and almost always square integrable. For most cases these solutions support bound states at zero energy. Some exceptional unbounded states are normalizable for non-zero angular momentum. Using a generalized definition, degeneracy of the nonrelativistic states is demonstrated and the associated degenerate observable is defined.

## I. INTRODUCTION

Exact solutions of the wave equation are very important because of the understanding of physics that can only be brought about by such solutions. Moreover, these solutions are valuable tools in checking and improving models and numerical methods being introduced for solving complicated physical problems; at least in some limiting cases. Exact solutions at $E = 0$ attracted attention lately [1-3] motivated in part by developments in supersymmetric quantum mechanics [4,5] and in the search for conditionally exactly and quasi exactly solvable problems [6,7]. From mathematical point of view these solutions are interesting since they form, by definition, quasi exactly solvable systems due to the fact that they are soluble only for $E = 0$. Moreover, and despite common intuition brought about by wide familiarity with the Coulomb problem, some of these solutions are square integrable and correspond to bound or unbound states [1,2,8]. There is another misconception about zero energy solutions where few of us, physicist, may inadvertently fall into and pass on to our students. It is the notion that these states are "dead" and "uninteresting" or "unimportant" because, presumably, there is nothing much happening. This might be true asymptotically where the kinetic energy vanishes, but in fact this cannot be farthest from the truth when we look closely to what is happening at finite separations form the center of interaction. There… the zero energy states are as good as any other, in fact they are full of kinetic energy as bound states; most have to go through potential barriers; some experience low energy capture, …etc. For the same reason it should also be understood that there is still a major distinction between the relativistic and nonrelativistic solutions despite the zero energy (strictly speaking, zero asymptotic kinetic energy). Furthermore, these solutions may turn out to be very valuable for zero energy limit calculations in various fields of physics. Such examples are in the study of loosely bound systems, as well as, in scattering length and coupling parameter calculations [9].

Exactly solvable problems in nonrelativistic quantum mechanics fall within distinct classes of shape invariant potentials [5,10-11] each belonging to a given dynamical symmetry group. Any potential in a given class can be mapped into another in the same class by a canonical transformation of the coordinates [11,12]. This transformation preserves the Schrödinger-like structure of the wave equation resulting in a correspondence map among the potential parameters, angular momentum, and energy. Using this parameter substitution and the bound states spectrum of the original potential one can easily and directly obtain the spectra of all other potentials in the class. Moreover, the eigenstates wavefunctions are obtained by simple transformations of those of the original potential. Well-known nonrelativistic shape invariant potentials can be divided in two classes based on the form of their eigenfunctions. In the first class, which includes the Coulomb, Oscillator and Mörse potentials, the wavefunctions are written in terms of the Confluent hypergeometric functions. Furthermore, these potentials belong to the dynamical symmetry group SO(2,1) [13]. In the second class the wavefunctions are written in terms of the hypergeometric functions. This class includes Rosen-Mörse, Eckart, Pöschl-Teller, and Scarf potentials. The algebraic expressions of these potentials and their properties are given in [5,10-11] and references therein.

Exact solutions of Dirac equation with potential interaction, on the other hand, are very rare. Since the original work of Dirac in the early part of last century up until 1989 only the relativistic Coulomb problem was solved exactly. In 1989, the relativistic extension of the oscillator problem (Dirac-Oscillator) was finally formulated and solved exactly by Moshinsky and Szczepaniak [14]. Recently, the Dirac-Mörse problem was also solved [15], its bound states spectrum and spinor wavefunctions were obtained. In general, the relativistic potential in Dirac equation has "even" and "odd" components, which are related by a "gauge fixing condition" [15,16]. In the search for relativistic extensions of solvable nonrelativistic problems, two important issues have to be observed. The first is that one should consider expressions for the radial potential components, which are simple linear combinations of basic functions such as $r$, $e^{-r}$, $\sin(r)$, $\tanh(r)$, etc. The use of the terms "simple" and "linear" is due to the fact that these are prominent features of the relativistic theory. Evidence of that can be found in many examples such as: (1) Dirac equation is linear in the derivative whereas Schrödinger equation is quadratic; (2) the Dirac-Oscillator potential [14] is linear in the coordinate while the oscillator potential is quadratic; (3) the Dirac-Mörse potential [15] is linear in the exponential (i.e. of the form $e^{-r}$) whereas the nonrelativistic Mörse potential is of the form $(1-e^{-r})^2$. The second issue is that a constraint should be imposed such that the two coupled first-order differential equations for the radial spinor components in Dirac equation result in a Schrödinger-like $2^{nd}$ order differential equation. This makes comparison with well-known exactly solvable nonrelativistic problems highly transparent. The imposed constraint turns out to be the gauge fixing condition that relates the even and odd components of the relativistic potential. Using this strategy, it was possible to obtain exact solutions for Dirac equation that are the relativistic extension of another class of shape invariant potentials [16], which includes Dirac-Pöschl-Teller, Dirac-Eckart, etc.

In this article we succeed in obtaining exact solutions of Dirac equation at rest mass energies for a relativistic potential whose odd component is of the form $\beta/r^{1+\beta}$, where $\beta$ is a real parameter and $-1 \neq \beta \neq -2$. This is accomplished by comparing this problem with an exactly solvable Schrödinger equation at zero nonrelativistic energy for a class of potentials of the form $r^{-2(\mu-\frac{1}{2})/(\mu+\frac{1}{2})} - D_{l,n} r^{-2\mu/(\mu+\frac{1}{2})}$, where $\mu$ is a real parameter such that $\mu \neq 0$ and $\mu \neq \pm\frac{1}{2}$, $n$ is a non-negative integer, $l$ is the angular momentum quantum number and $D_{l,n}$ is a positive parameter that depends linearly on $l$ and $n$. The latter is solved by performing a point canonical transformation of the well-known exactly solvable problem of the three dimensional harmonic oscillator. This is accomplished in section II utilizing the potential algebra approach of SO(2,1) [13,17]. The wavefunction solutions are written in terms of the confluent hypergeometric functions. They are square integrable except when $l = 0$ and $\mu < -\frac{1}{2}$. For most cases, depending on the values of the physical parameters, these solutions support bound states at zero energy. We also found unbounded states, which are normalizable for non-zero angular momentum. This result supports earlier findings by other authors [1,18] where these unbounded normalizable states are termed "exceptional". The issue of degeneracy of the solution space is addressed and illustrated within the generalized definition that "*states with different parameter values produce the same value of a given observable*". The observable, which is associated with this degeneracy, is defined. In section III, the relativistic problem is formulated and a correspondence map between the physical parameters of the relativistic and nonrelativistic problem is found resulting in constraints for obtaining the exact relativistic solution. It turns out that the integer index $n$ can ONLY take the value zero. Moreover, and in contrast with the nonrelativistic theory, the coupling constant of the relativistic potential is independent of the angular momentum. The correspondence map gives, as well, the spinor wavefunction solutions via a simple transformation of those of the nonrelativistic problem.

## II. THE NONRELATIVISTIC PROBLEM

Starting with the exactly solvable problem of the 3D oscillator in the Appendix one can obtain solutions of other problems belonging to the same class as that of the oscillator using "point canonical transformations" [11,12]. This is accomplished by performing coordinate transformations of equation (A.5) that preserves the Schrödinger–like structure of the equation (i.e., the first order derivative terms vanish). It should also leave the form of the function $f(x)$, as given by equation (A.3), invariant (shape-invariant). This transformation is written as

$$r = g(x) \quad , \quad \psi(r) = h(x)\Phi(x) \qquad (2.1)$$

The Schrödinger–like constraint requires that $h = \sqrt{dg/dx}$. Then, the differential equation reads

$$\left[\frac{d^2}{dr^2} + f(r)\right]\psi(r) = (g')^{-3/2}\left\{\frac{d^2}{dx^2} + \frac{1}{2}\left[\frac{g'''}{g'} - \frac{3}{2}\left(\frac{g''}{g'}\right)^2\right] + (g')^2 f(g)\right\}\Phi = 0$$

where the prime symbol stands for the derivative with respect to the coordinate $x$. This equation is equivalent to (A.5) if we write:

$$f(r) = (g')^{-2}\left[-\frac{4\gamma(\gamma+1)+3/4}{x^2} - \lambda^4 x^2 + 4\lambda^2(\gamma+n+1) - \frac{1}{2}\frac{g'''}{g'} + \frac{3}{4}\left(\frac{g''}{g'}\right)^2\right] \qquad (2.2)$$

$$\equiv -\frac{l(l+1)}{r^2} - 2V(r) + 2E$$

The physical parameters $\gamma$ and $\lambda$ are defined in the Appendix. The choice $g(x) = x^2$ gives the exact solution for the nonrelativistic Coulomb problem. On the other hand, taking $g(x) = -\ln(x)$ solves the S-wave ($l = 0$) Mörse problem. As for the present problem, our choice will be $g(x) = x^{2\mu+1}$, where $\mu$ is a real parameter and $\mu \neq 0, \pm\frac{1}{2}$. The three dismissed values of $\mu$ correspond to the Oscillator, Coulomb, and Mörse problems, respectively. Substituting this transformation function in (2.2) gives the following physical parameters

$$E = 0$$
$$\gamma = -1/2 + (l + \tfrac{1}{2})|\mu + \tfrac{1}{2}|$$

and results in the following expression for the potential function:

$$V(r) = \left(\frac{\lambda}{2\mu+1}\right)^2 \left[\frac{\lambda^2/2}{r^{2(\mu-\frac{1}{2})/(\mu+\frac{1}{2})}} - \frac{2n+1+(2l+1)|\mu+\frac{1}{2}|}{r^{2\mu/(\mu+\frac{1}{2})}}\right] \quad ; n = 0,1,2,\ldots \quad (2.3)$$

which is a two-term power-law potential. For a given parameter $\mu$ and angular momentum $l$ there are infinitely many potentials parameterized by the integer $n$ that solve Schrödinger equation exactly at zero energy. Note that the first term in the potential is repulsive while the second, which has a coupling constant that depends on the angular momentum $l$ and integer $n$, is attractive. Note also that in the limit both terms go to $1/r^2$ as $\mu$ goes to $\pm\infty$. Figure (1) shows how the two exponents of $r$ change with the parameter $\mu$. The graph clearly shows that there are only three values of $\mu$ (0 and $\pm\frac{1}{2}$) where one of the two exponents either vanishes or becomes undetermined resulting in a constant term that can be interpreted as energy. These cases correspond to the Oscillator, Coulomb and Mörse problems, respectively.

Using the transformation in (2.1) with this $g(x)$ and the oscillator wavefunction in equation (A.6) of the Appendix we can write down the wavefunction for this zero-energy problem, depending on whether $\mu$ is greater than or less than $-\frac{1}{2}$, as follows:

$$\psi_n^l(r) = a_n \exp\left[-(\lambda^2/2)r^{1/(\mu+\frac{1}{2})}\right] L_n^{(2l+1)|\mu+\frac{1}{2}|}\left(\lambda^2 r^{1/(\mu+\frac{1}{2})}\right) \times \begin{cases} \left(\lambda^{2\mu+1}r\right)^{l+1} & ;\mu > -\frac{1}{2} \\ \left(\lambda^{2\mu+1}r\right)^{-l} & ;\mu < -\frac{1}{2} \end{cases} \quad (2.4)$$

For square integrable wavefunctions $a_n$ is the normalization constant.

It is evident from the expression of the potential in (2.3) that there are special states that solve the wave equation for the same potential. This occurs if, for a given $\mu$, the parameters $n$ and $l$ of these states satisfy the condition

$$2n + 1 + (2l+1)|\mu + \tfrac{1}{2}| = \text{a real constant} \equiv \Omega$$

This is in fact a form of (accidental) degeneracy in the general sense of its definition: "*states with different parameter values produce the same value of a given observable*". For example, in the Coulomb problem of the Hydrogen atom, states of different angular momenta and indices $\{l,n\}$ produce the same principal quantum number $N_p$ which is equal to $l + n + 1$ (i.e., they produce the same energy level which is proportional to $-1/N_p^2$). In our present problem, the observable is some real function of the physical constant $\Omega$. Degeneracy almost always gives a constraint among the physical parameters similar to that in the above relation. In the example of the Hydrogen atom, the constraint is $l + n + 1 =$ constant. As a simple illustration of degeneracy for our problem, we consider the case where $\mu = 3/2$. Then all odd values of $\Omega$ will produce degeneracy of an order equals to the nearest integer to $\Omega/4$. Specifically, if we take $\Omega = 11$ then states with the following set of $(l,n)$-pairs are degenerate: (0,4), (1,2), and (2,0). For $\Omega = 13$, the three degenerate states are those with the following $(l,n)$-pairs: (0,5), (1,3), and (2,1).

Now, we will address the issues of boundedness and normalizability and consider the two cases, where $\mu > -\frac{1}{2}$ or $\mu < -\frac{1}{2}$, separately:

**Case (1): $\mu > -\frac{1}{2}$**

We write $\mu = -\frac{1}{2} + \beta^{-1}$, where $1 \neq \beta > 0$ and $\beta \neq 2$. In this case the effective potential [the sum of the angular momentum component and $V(r)$ in (2.3)] takes the following form

$$V_{eff} \sim \frac{1}{r^2}\left[ l(l+1) + \frac{\beta^2 \lambda^4}{4} r^{2\beta} - \frac{\beta^2 \lambda^2}{2}\left(2n + 1 + \frac{2l+1}{\beta}\right) r^{\beta}\right] \quad (2.5)$$

- For $1 > \beta > 0$ (i.e. $\mu > \frac{1}{2}$) the tail of the potential approaches zero from above as $r \to \infty$, that is the $\lim_{r \to \infty} V_{eff} = +0$. Moreover, for $l = 0$ the $\lim_{r \to 0} V_{eff} = -\infty$, which implies that the potential function rises upward crossing the zero energy line toward a maximum then drops gradually to zero from above. This behavior is illustrated graphically in Figure (2). Therefore, this case supports bound states at zero energy. However, for $l > 0$ the $\lim_{r \to 0} V_{eff} = +\infty$ and there exist two situations, depending on the range of values of the physical parameters, one can support bound states for $E = 0$ where $V_{eff}$ has a local minimum and maximum, while the other doesn't. The necessary (but not sufficient) condition for bound states in this case is:

$$n > 2\frac{\sqrt{l(l+1)(1-\beta)}}{\beta(2-\beta)} - \frac{l+\frac{1}{2}}{\beta} - \frac{1}{2} \quad (2.6)$$

These two situations are also shown on the same Figure (2).

- Now, for $2 \neq \beta > 1$ (i.e. $\frac{1}{2} > \mu > -\frac{1}{2}$, but $\mu \neq 0$), the $\lim_{r \to \infty} V_{eff} = +\infty$ and the potential is strongly confining. For $l = 0$ the $\lim_{r \to 0} V_{eff}$ either vanishes (for $\beta > 2$) or goes to $-\infty$ (for $2 > \beta > 1$), and both cases result in bounded states. For $l > 0$ the $\lim_{r \to 0} V_{eff} = +\infty$ and the zero energy state is also bounded. Figure (3) shows these behaviors graphically.

- The wavefunction in this case and for all positive values of $\beta$ takes the form $\psi \sim r^{l+1} e^{-r^{\beta}} L_n^{(2l+1)/\beta}(2r^{\beta})$, which has always a finite norm (i.e. square integrable). This implies that even the unbounded zero energy state (in the case where $1 > \beta > 0$ and $l > 0$) is normalizable. For this reason, these are called "exceptional" states.

**Case (2): $\mu < -\frac{1}{2}$**

We write $\mu = -\frac{1}{2} - \beta^{-1}$, where $\beta > 0$. In this case the effective potential function takes the following form

$$V_{eff} \sim \frac{1}{r^2}\left[ l(l+1) + \frac{\beta^2 \lambda^4}{4} \frac{1}{r^{2\beta}} - \frac{\beta^2 \lambda^2}{2}\left(2n + 1 + \frac{2l+1}{\beta}\right) \frac{1}{r^{\beta}}\right] \quad (2.7)$$

- The $\lim_{r \to 0} V_{\mathit{eff}} = +\infty$. Moreover, for $l = 0$ the $\lim_{r \to \infty} V_{\mathit{eff}} = -0$, which implies no bound states at $E = 0$ as shown in Figure (4). However, for $l > 0$ the $\lim_{r \to \infty} V_{\mathit{eff}} = +0$ and again there are two parameter-dependent situations, one can support bound states at $E = 0$ while the other doesn't as illustrated in Figure (4). The necessary (but not sufficient) condition for bound states in this case is:

$$n > 2\frac{\sqrt{l(l+1)(1+\beta)}}{\beta(2+\beta)} - \frac{l+\frac{1}{2}}{\beta} - \frac{1}{2} \tag{2.8}$$

- The wavefunction takes the form $\psi \sim r^{-l} e^{-r^{-\beta}} L_n^{(2l+1)/\beta}(2r^{-\beta})$. To see whether this wavefunction is normalizable or not, we make the change of variable $r \to \rho = r^{-\beta}$ giving the following norm

$$\|\psi_n^l\| = \int_0^\infty (\psi_n^l)^2 \, dr \sim \int_0^\infty \rho^{-1+(2l-1)/\beta} e^{-2\rho} \left[ L_n^{(2l+1)/\beta}(2\rho) \right]^2 d\rho$$

This shows that the wave function is normalizable for all $l > 0$. Therefore, the unbounded state for $l = 0$ is not normalizable, while that for $l > 0$, which is exceptional, is normalizable.

Normalizable unbounded states at zero energy have been reported elsewhere [1,18]. In the present problem, this happens in some cases as shown above and only for $l > 0$. A summary of normalizability and boundedness is given in Table I.

An interesting special case is when $\mu = 3/2$ giving

$$V(r) = \frac{Z}{r} - \frac{\sqrt{Z/2}(2l+n+3/2)}{r^{3/2}}$$

$$\psi_n^l(r) = a_n (Zr)^{l+1} \exp\left(-2\sqrt{2Zr}\right) L_n^{4l+2}\left(4\sqrt{2Zr}\right)$$

where $Z > 0$. It is a repulsive Coulomb problem plus an attractive angular-momentum-dependent potential component. $\mu = 1/6$ gives a solution, at zero energy, for the problem with linear potential plus an angular-momentum-dependent component of the form $1/\sqrt{r}$. Moreover, $\mu = -3/4$ gives solution to the problem with $1/r^4$ potential plus an $(l,n)$-dependent component of the form $1/r^6$.

III. THE RELATIVISTIC PROBLEM

Using gauge invariance of the theory of interaction of charged particles with the electromagnetic field, it has been shown [15,16] that the 2-component radial Dirac equation for a charged spinor in spherically symmetric electromagnetic 4-potential can generally be written as

$$\begin{pmatrix} 1+\alpha^2 \mathcal{V}(r) & \alpha\left(\frac{\kappa}{r}+\mathcal{W}(r)-\frac{d}{dr}\right) \\ \alpha\left(\frac{\kappa}{r}+\mathcal{W}(r)+\frac{d}{dr}\right) & -1+\alpha^2 \mathcal{V}(r) \end{pmatrix} \begin{pmatrix} \phi(r) \\ \theta(r) \end{pmatrix} = \varepsilon \begin{pmatrix} \phi(r) \\ \theta(r) \end{pmatrix} \tag{3.1}$$

where $\alpha$ is the fine structure constant, $\varepsilon$ is the relativistic energy and $\kappa$ is the spin-orbit coupling parameter defined as $\kappa = \pm (j + \frac{1}{2})$ for $l = j \pm \frac{1}{2}$. The real radial functions $\mathcal{V}(r)$ and $\mathcal{W}(r)$ are the even and odd parts of the relativistic potential, respectively.

The relativistic problem at rest mass energies (i.e., $\varepsilon = 1$) is formulated by taking the even component of the relativistic potential $\mathcal{V}(r) = 0$ and considering the following expression for the odd component

$$\mathcal{W}(r) = \frac{A}{r^{(v-1/2)/(v+1/2)}} \tag{3.2}$$

where $A$ is a real coupling constants and the real parameter $v \neq -\frac{1}{2}$. The expression for $\mathcal{W}(r)$ above was chosen by following the strategy that was stated in the introduction with an eye on the form of the nonrelativistic potential in equation (2.3). The resulting second order differential equation for the upper spinor component, which is obtained from Dirac equation (3.1) with $\mathcal{V}(r) = 0$ and $\varepsilon = 1$, reads:

$$\left[ -\frac{d^2}{dr^2} + \frac{\kappa(\kappa+1)}{r^2} + \mathcal{W}^2 - \frac{d\mathcal{W}}{dr} + 2\kappa \frac{\mathcal{W}}{r} \right] \phi(r) = 0 \tag{3.3}$$

Using the expression of $\mathcal{W}(r)$ in (3.2) gives the following Schrödinger-like wave equation

$$\left\{ -\frac{d^2}{dr^2} + \frac{\kappa(\kappa+1)}{r^2} + \frac{A^2}{r^{2(v-1/2)/(v+1/2)}} + \frac{A[2\kappa + (v-1/2)/(v+1/2)]}{r^{2v/(v+1/2)}} \right\} \phi(r) = 0$$

Comparison of this equation with Schrödinger equation for the nonrelativistic problem, whose potential is given by (2.3), results in the following correspondence map among the parameters of the two problems:

$$v = \mu$$
$$\kappa = l \quad \text{or} \quad \kappa = -l - 1 \tag{3.4}$$
$$A = \lambda^2/(2v+1)$$

However, compatibility among these relations is maintained ONLY for $n = 0$ and in either one of two possible alternatives:

(i) $n = 0$, $\kappa = l$, and $v < -\frac{1}{2}$,

(ii) $n = 0$, $\kappa = -l - 1$, and $v > -\frac{1}{2}$,

If we write $v = -\frac{1}{2} + \beta^{-1}$, where $\beta$ is real, finite and $1 \neq \beta \neq 2$, then the odd component of the relativistic potential can be written as

$$\mathcal{W}(r) = \frac{\lambda^2 \beta/2}{r^{1-\beta}}$$

Hence, the relativistic potential is independent of the angular momentum, in contrast with the nonrelativistic problem. However, the corresponding potential obtained by reduction to the Schrödinger-like wave equation (3.3) reads

$$V(r) = \frac{\lambda^2 \beta}{4} \left[ \frac{\lambda^2 \beta/2}{r^{2(1-\beta)}} + \frac{2\kappa - \beta + 1}{r^{2-\beta}} \right]$$

The nonrelativistic wavefunction given by (2.4) is transformed into the following upper spinor component:

$$\phi_l^\beta(r) = C_l \left( \lambda^{2/\beta} r \right)^{-\kappa} \exp\left( -\frac{\lambda^2}{2} r^\beta \right)$$

It is normalizable except when $l = 0$ and $\beta < 0$. For normalizable states, $C_l$ is the normalization constant:

$$C_l = \lambda^{1/\beta} \sqrt{|\beta|/\Gamma\left(\frac{1-2\kappa}{\beta}\right)}$$

The lower spinor component is obtained from the upper using Dirac equation (3.1) with $\mathcal{V}(r) = 0$ and $\varepsilon = 1$. That is,

$$\theta(r) = \frac{\alpha}{2}\left(\mathcal{W} + \frac{\kappa}{r} + \frac{d}{dr}\right)\phi(r)$$

Substituting the expression of $\mathcal{W}(r)$ above we find that the lower component vanishes, as expected in the case of zero energy solutions.

## IV. ADDENDUM

An interesting paper by C M Bender and Q Wang [22] appeared after our article was submitted for publication to the Physical Review A. It addresses a problem that relates closely to our nonrelativistic case and deserves a comment. In the cited paper, the authors obtain exact solutions for the 2$^{nd}$ order differential equation

$$\left[-\frac{d^2}{dx^2} + x^{2(N+1)} - Ex^N\right]\psi(x) = 0 \qquad (4.1)$$

on the real line $-\infty < x < \infty$ for $N = -1, 0, 1, 2, \ldots$ and for real "eigenvalue" parameter $E$.

In the semi-infinite interval ($x \geq 0$) this is a special case of the nonrelativistic problem in section II. The following parameter substitution reproduces equation (4.1):

$l = 0$

$\mu = -\tfrac{1}{2} + (N+2)^{-1}$

$\lambda^2 = 2/|N+2|$

The dismissed values of $\mu$ in section II correspond to $N = -1$ and $N = 0$ which refers to the S-wave Coulomb and S-wave oscillator problems, respectively. The restriction in the problem of the cited paper, as opposed to ours, is three-fold: (1) the angular momentum vanishes, (2) the powers of $x$ are discrete, and (3) the inverse power-law is limited to only $-1$. Comparing equation (4.1) with the wave equation in section II, whose potential is given by (2.3), results straightforwardly in the following discrete values for $E$:

$$E_n = (2n+1)|N+2| + 1 = |N+2|\Omega \qquad ; n = 0, 1, 2, \ldots$$

Moreover, equation (2.4) gives the following eigenfunctions:

$$\psi_n(r) = a_n \exp\left[-(\lambda^2/2)r^{N+2}\right] L_n^{1/|N+2|}\left(\lambda^2 r^{N+2}\right) \times \begin{cases} \lambda^{2/(N+2)} r & ; N > -2 \\ 1 & ; N < -2 \end{cases}$$

In the cited paper the authors give much more detailed description of the eigenfunctions, which are also written in terms of the confluent hypergeometric functions. They consider the case for even and odd $N$ separately. In addition, they also write the eigenfunctions using orthogonal polynomials that have interesting mathematical properties.

ACKNOWLEDGMENTS

The financial support of ABU is highly appreciated and acknowledged. The author is grateful to Dr. H. A. Yamani for very fruitful discussions. Many thanks go to Dr. M. S. Abdelmonem for the support in literature survey.

## APPENDIX: SO(2,1) POTENTIAL ALGEBRA FOR EXACTLY SOLVABLE NONRELATIVISTIC PROBLEMS

In the potential algebra approach [13,17], the dynamical symmetry of the physical problem is exploited by studying the representations of its spectrum generating algebra. Realizations of the generators of such algebra by differential operators facilitate study of the properties of the wave equation and its solutions. Examples of these are: SO(2,1) algebra for solving some three-parameter problems such as the harmonic oscillator [13,17]; the conformal group algebra SO(4,2) in the study of the extended symmetry of the Coulomb and Kepler problems [19]; the algebra SO(3,2) in the investigation of field theories in anti-de Sitter space like the theory of the Dirac supermultiplet (the singletons) [20]. For our present problem, a brief overview of the SO(2,1) Lie algebra, its discrete representations, and operator realizations is called for.

SO(2,1) is a three dimensional Lie algebra whose generators satisfy the commutation relations $[L_3, L_\pm] = \pm L_\pm$ and $[L_+, L_-] = -L_3$, where $L_3^\dagger = L_3$ and $L_\pm^\dagger = L_\mp$. It is an algebra of rank one and has one Casimir invariant operator. This operator, which commutes with $L_3$ and $L_\pm$, is $L^2 = L_3(L_3 \pm 1) - 2L_\mp L_\pm$. Among the four operators $L^2$, $L_3$ and $L_\pm$ there is a maximum of two commuting, one of which is $L^2$. To obtain the discrete representation, we choose the compact operator $L_3$ to commute with $L^2$ rather than the non-compact $L_\pm$. Therefore, $L_3$ shares the same eigenvectors with $L^2$. Elements of this representation are labeled with two parameters corresponding to the eigenvalues of $L^2$ and $L_3$. In fact there are three discrete unitary representations of SO(2,1) [13]. Two of them are bounded at one end; one with a lower bound and the other with an upper bound. The third is not bounded. Physically, we are interested in the one that is bounded by a ground state from below. It is parameterized by a real constant $\gamma \geq -1/2$ and denoted as $D^+(\gamma)$. The action of the operators of the algebra on the basis $|\gamma, n\rangle$ is as follows:

$$L^2 |\gamma, n\rangle = \gamma(\gamma+1)|\gamma, n\rangle$$
$$L_3 |\gamma, n\rangle = (\gamma+n+1)|\gamma, n\rangle$$
$$L_+ |\gamma, n\rangle = \frac{1}{\sqrt{2}}\sqrt{(n+1)(n+2\gamma+2)}|\gamma, n+1\rangle \quad ; n = 0,1,2,\ldots \quad \text{(A.1)}$$
$$L_- |\gamma, n\rangle = \frac{1}{\sqrt{2}}\sqrt{n(n+2\gamma+1)}|\gamma, n-1\rangle$$

Realization of the generators of SO(2,1) algebra in terms of differential operators in one variable is of great importance since it is intended to solve the physically interesting second order differential wave equations of the form:

$$\left[\frac{d^2}{dx^2} + f(x)\right]\Phi(x) = 0 \quad \text{(A.2)}$$

where
$$f(x) = -\frac{l(l+1)}{x^2} - 2V(x) + 2E \quad (A.3)$$

$l$ is the angular momentum quantum number, $E$ is the energy, and $V(x)$ is a real potential function. The most general form of the three generators $L_3$ and $L_\pm$ whose linear combination gives the second order differential operator in (A.2) are:

$$L_3 = \frac{d^2}{dx^2} + b_3(x)\frac{d}{dx} + a_3(x)$$

$$L_\pm = \frac{1}{\sqrt{2}}\left[\frac{d^2}{dx^2} + b_\pm(x)\frac{d}{dx} + a_\pm(x)\right]$$

Hermiticity and the fact that in the space of $L^2(0,\infty)$ functions, $\overline{(d/dx)}^\dagger = -\overline{(d/dx)}$ give $b_- = -b_+$, $a_- = a_+ - db_+/dx$, and $b_3 = 0$. Moreover, applying the commutation relations and performing some manipulations, we arrive at the following:

$$L_3 = \frac{d^2}{dx^2} + \frac{\eta}{x^2} - \frac{x^2}{16}$$

$$L_\pm = \frac{1}{\sqrt{2}}\left[\frac{d^2}{dx^2} + \frac{\eta}{x^2} + \frac{x^2}{16} \pm \frac{1}{2}\left(x\frac{d}{dx} + \frac{1}{2}\right)\right]$$

where $\eta$ is a real constant parameters. As a result of this realization, one finds that

$$L^2 = -\frac{1}{4}\left(\eta + \frac{3}{4}\right) \equiv \gamma(\gamma+1)$$

Thus giving $\eta = -4\gamma(\gamma+1) - 3/4$, and $\eta \leq ¼$. The Hamiltonian of a system whose spectrum is generated by the representation of SO(2,1), must be an element in the linear vector space spanned by its generators. That is we can expand such a Hamiltonian, $H$, as a linear combination of these generators and as follows:

$$-2H = \sqrt{2}\tau_+ L_+ + \sqrt{2}\tau_- L_- + \tau_3 L_3$$

where $\tau_\pm^* = \tau_\mp$, $\tau_3^* = \tau_3$ are constant parameters. Using the realization of $L_3$ and $L_\pm$ obtained above, we can write this as

$$-2H = (\tau_+ + \tau_- + \tau_3)\frac{d^2}{dx^2} + \frac{1}{2}(\tau_+ - \tau_-)x\frac{d}{dx} + (\tau_+ + \tau_- + \tau_3)\frac{\eta}{x^2}$$

$$+ (\tau_+ + \tau_- - \tau_3)\frac{x^2}{16} + \frac{1}{4}(\tau_+ - \tau_-)$$

Therefore, $\tau_+ + \tau_- + \tau_3 = 1$. Moreover, to obtain a Schrödinger-like equation, the first order derivative has to be eliminated which requires that $\tau_+ = \tau_- = (1-\tau_3)/2$ giving:

$$-2H = (1-\tau_3)L_1 + \tau_3 L_3 = \frac{d^2}{dx^2} + \frac{\eta}{x^2} + \frac{1-2\tau_3}{16}x^2$$

where $L_\pm = (L_1 \pm iL_2)/\sqrt{2}$. Schrödinger wave equation is the eigenvalue equation $(H-E)\Phi = 0$ which can now be written as

$$[(1-\tau_3)L_1 + \tau_3 L_3 - \tau_0]\Phi = \left(\frac{d^2}{dx^2} + \frac{\eta}{x^2} + \frac{1-2\tau_3}{16}x^2 - \tau_0\right)\Phi(x) = 0 \quad (A.4)$$

where $\tau_0$ is the real eigenvalue which is equal to $2E$. Using only the commutation relations of SO(2,1) we can perform a unitary transformation called the "tilting transformation":

$$e^{i\zeta L_2}(L_3 \pm L_1)e^{-i\zeta L_2} = e^{\mp\zeta}(L_3 \pm L_1)$$

where $\zeta$ is a real constant parameter. This transforms Schrödinger equation (A.4) to

$$\left\{\left[1-(2\tau_3-1)\ e^{2\zeta}\right]L_1 + \left[1+(2\tau_3-1)\ e^{2\zeta}\right]L_3 - 2\tau_0 e^{\zeta}\right\}\Phi = 0$$

To obtain the discrete representation, we choose the "tilting angle" $\zeta$ such that the coefficient of $L_1$ vanishes. We should note, however, that the range of possible values of $\zeta$ is restricted by the value of $\tau_3$ in the problem. For bound states $\tau_3 > 1/2$, while, for scattering states (the continuum) $\tau_3 < 1/2$. However, presently we require that

$$(2\tau_3 - 1)e^{2\zeta} = 1 \quad \Rightarrow \quad \tau_3 > 1/2$$

Thus, giving

$$(L_3 - \tau_0 e^{\zeta})\Phi = 0 \quad \Rightarrow \quad L_3\Phi = \frac{\tau_0}{\sqrt{2\tau_3 - 1}}\Phi$$

Using the spectrum of $L_3$ in equation (A.1), we can write

$$\frac{\tau_0}{\sqrt{2\tau_3 - 1}} = \gamma + n + 1 \quad ; \quad n = 0, 1, 2, ...$$

Now, if we define $(2\tau_3 - 1)/16 \equiv \lambda^4$, then we can finally write the differential equation (A.4) as follows

$$\left[\frac{d^2}{dx^2} - \frac{4\gamma(\gamma+1)+3/4}{x^2} - \lambda^4 x^2 + 4\lambda^2(\gamma + n + 1)\right]\Phi_n^\gamma(x) = 0 \tag{A.5}$$

where $\Phi_n^\gamma(x) \equiv \langle x|\gamma, n\rangle$. The normalized solution of this differential equation can be written as [21]:

$$\Phi_n^\gamma(x) = \sqrt{\frac{2\lambda\ \Gamma(n+1)}{\Gamma(2\gamma+n+2)}}(\lambda x)^{2\gamma+3/2}\ e^{-\lambda^2 x^2/2}L_n^{2\gamma+1}(\lambda^2 x^2) \tag{A.6}$$

$\Gamma$ is the gamma function and $L_n^\nu(x)$ are the Laguerre polynomials. Equation (A.5) is, evidently, Schrödinger equation for the three-dimensional harmonic oscillator, where $\lambda$ is the oscillator strength parameter and $\gamma = (l - \frac{1}{2})/2$. It also gives the spectrum of the bound states whose corresponding normalized wavefunctions are those given by equation (A.6). Canonical transformations of the coordinates, similar to those in (2.1) of section II, map this differential equation into other solvable problems belonging to the same class as that of the oscillator. These transformations produce a correspondence map among the physical parameters of the two problems leading to the new spectrum and eigenstates wavefunctions.

FIGURES CAPTION

FIG. 1
The exponents of the two power-law potential terms in equation (2.3) as a function of the parameter $\mu$. In the limit both terms go to $1/r^2$ as $\mu$ goes to $\pm\infty$. It also shows that there are only three values of $\mu$ (0 and $\pm\frac{1}{2}$) where one of the two exponents either vanish or becomes undetermined giving a constant that can be interpreted as energy. These cases correspond to the Oscillator, Coulomb and Mörse problems, respectively

FIG. 2
The general behavior of the effective potential in equation (2.5) where $\mu > \frac{1}{2}$ for the cases: (a) $l = 0$ resulting in bound states for $E = 0$; (b) $l > 0$ with condition (2.6) being satisfied and sufficient resulting in bound states at $E = 0$; (c) $l > 0$ but the condition (2.6) is not satisfied resulting in unbound but normalizable exceptional states.

FIG. 3
Illustration of the general behavior of the effective potential in equation (2.5) where $\frac{1}{2} > (\mu \neq 0) > -\frac{1}{2}$ for the cases: (a) $l = 0$ and $\frac{1}{2} > \mu > 0$; (b) $l = 0$ and $0 > \mu > -\frac{1}{2}$; (c) $l > 0$. All of these cases result in bound states for $E = 0$.

FIG. 4
The general behavior of the effective potential in equation (2.7) where $\mu < -\frac{1}{2}$ for the cases: (a) $l = 0$ which results in unbound unnormalizable states at $E = 0$; (b) $l > 0$ and the condition (2.8) being satisfied and sufficient resulting in bound states at $E = 0$; (c) $l > 0$ but the condition (2.8) is not satisfied resulting in unbound, however, normalizable exceptional states for $E = 0$.

TABLE CAPTION

The table summarizes normalizability and boundedness of the nonrelativistic zero energy solutions. In the Bounded column the equation numbers in parentheses are for the necessary condition of boundedness.

TABLE I

| $\mu$ | $l$ | Bounded | Normalizable |
|---|---|---|---|
| $\mu < -\tfrac{1}{2}$ | $l = 0$ | ✘ | ✘ |
|  | $l > 0$ | ✔ (2.8) | ✔ |
|  | $l > 0$ | ✘ (2.8) | ✔ |
| $\mu > \tfrac{1}{2}$ | $l = 0$ | ✔ | ✔ |
|  | $l > 0$ | ✔ (2.6) | ✔ |
|  | $l > 0$ | ✘ (2.6) | ✔ |
| $\tfrac{1}{2} > \mu > -\tfrac{1}{2}$ | $l = 0$ | ✔ | ✔ |
| $\mu \neq 0$ | $l > 0$ | ✔ | ✔ |

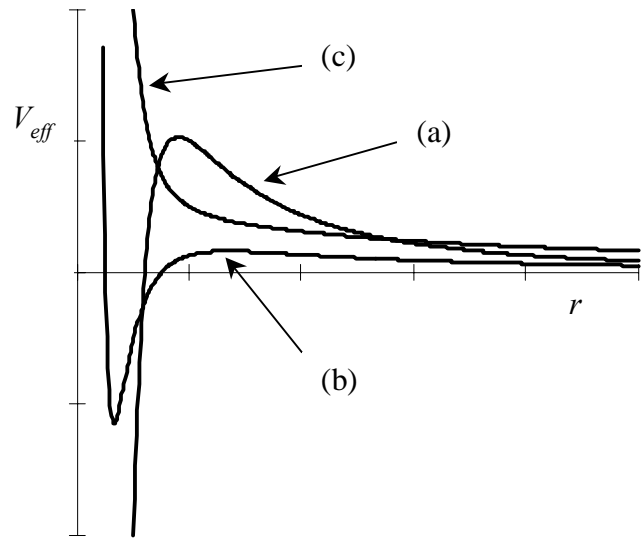

Figure (2)

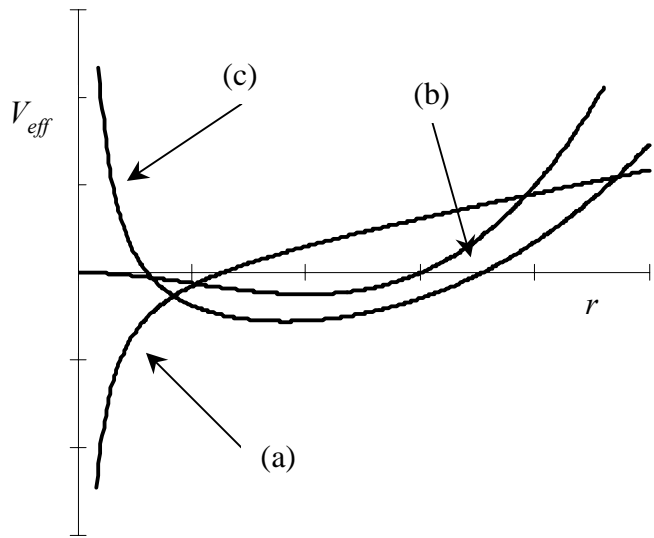

Figure (3)

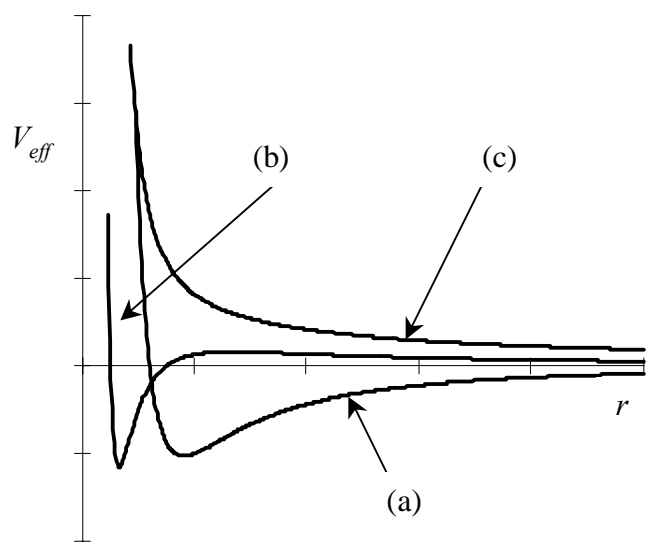

Figure (4)

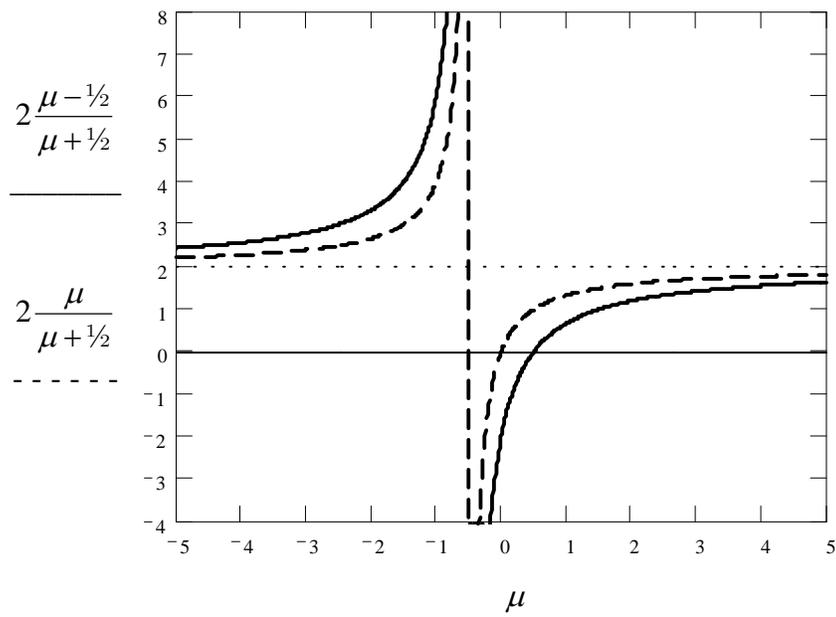

Figure (1)